# Plastics and Composite Materials

*A. T. Pérez Fontenla*
CERN, Geneva, Switzerland

**Abstract**
Polymers and composite materials play an essential role in accelerator and detectors technology, with varying roles that range from electrical insulation and structural support to thermal management. This paper provides a general review of their key properties and classifications, including behaviour under demanding service conditions such as cryogenic operation and high radiation exposure. The paper addresses polymeric materials—their mechanical, thermal, and viscoelastic behaviour, and the effects of crystallinity and additives—alongside composite families, focusing on the characteristics of the matrix and the types of reinforcement. CERN case studies illustrate how both polymers and composites present opportunities and challenges in material selection. Examples include adhesives and structural composites for detectors, reinforced alloys for collimators, and insulation for $Nb_3Sn$ superconducting magnets, all emphasising the need to optimise material properties and interfaces to ensure the long-term reliability of components in accelerator facilities.

## 1 Polymeric Materials

### 1.1 Introduction

Polymers, commonly referred to as plastics, re found throughout daily life, from clothing and food packaging to transport and protective equipment. Beyond these everyday uses, they are integral to advanced technologies, enabling applications from the space programme and bullet-resistant vests to biomedical implants. Their widespread adoption arises from a unique combination of properties: ease of fabrication, durability, versatility, corrosion resistance, low density with resilience, and effective thermal, electrical, and acoustic insulation. Moreover, polymers can be produced in an extraordinary variety of forms—from fibres, sheets, and foams to complex moulded components—making them indispensable across diverse fields.

### 1.2 Definition and classification

A polymer is a macromolecule composed of repeating structural units (monomers) linked by covalent bonds. The term derives from the Greek words *poly* (“many”) and *meros* (“parts” or “units”). In everyday language, *plastics* usually refer to organic polymers—natural or synthetic—often combined with additives to enhance performance or processability.

Polymers may be natural, such as cellulose, cotton, and latex, or synthetic, produced from small organic molecules such as ethylene or styrene. Compared to their natural counterparts, synthetic polymers generally offer superior properties at lower production costs. The chemical modification of natural polymers began in the 19th century with the development of celluloid from cellulose and vulcanized rubber from latex with the first fully synthetic polymer, Bakelite, produced in 1907. Large-scale industrial production of synthetic polymers did not occur until Second World War, when natural resources became scarce and the development of alternatives accelerated.

Most polymers are carbon-based, and the polymerisation consists of chemically bonding monomers into chains. A classic example is polyethylene (PE), derived from the simple monomer ethylene ($C_2H_4$), a gas at ambient conditions with a carbon–carbon double bond, as shown in Fig. 1(a). Under suitable conditions, a catalyst initiates polymerisation, leading to the successive addition of monomer units into a long chain. While often represented schematically as a linear chain, the actual molecular geometry reflects a tetrahedral bond angle (~109°), producing a zigzag backbone conformation in three dimensions as illustrated in Fig. 1(b).

Polymer chains bend, coil, and entangle, giving polymers their characteristic elasticity, especially in rubbers. Their mechanical and thermal behaviour depends on the ease of chain rotation. As shown in Fig. 1(a)–(c), PE has a simple backbone that allows free rotation, producing a flexible, ductile material often used in food packaging films while polystyrene (PS) presents a bulky side groups (phenyl ring) that restrict rotation, making the polymer stiffer and more brittle. These examples illustrate how small differences in molecular structure can dictate macroscopic performance, and why chain conformation is critical for predicting mechanical and thermal behaviour [1–2].

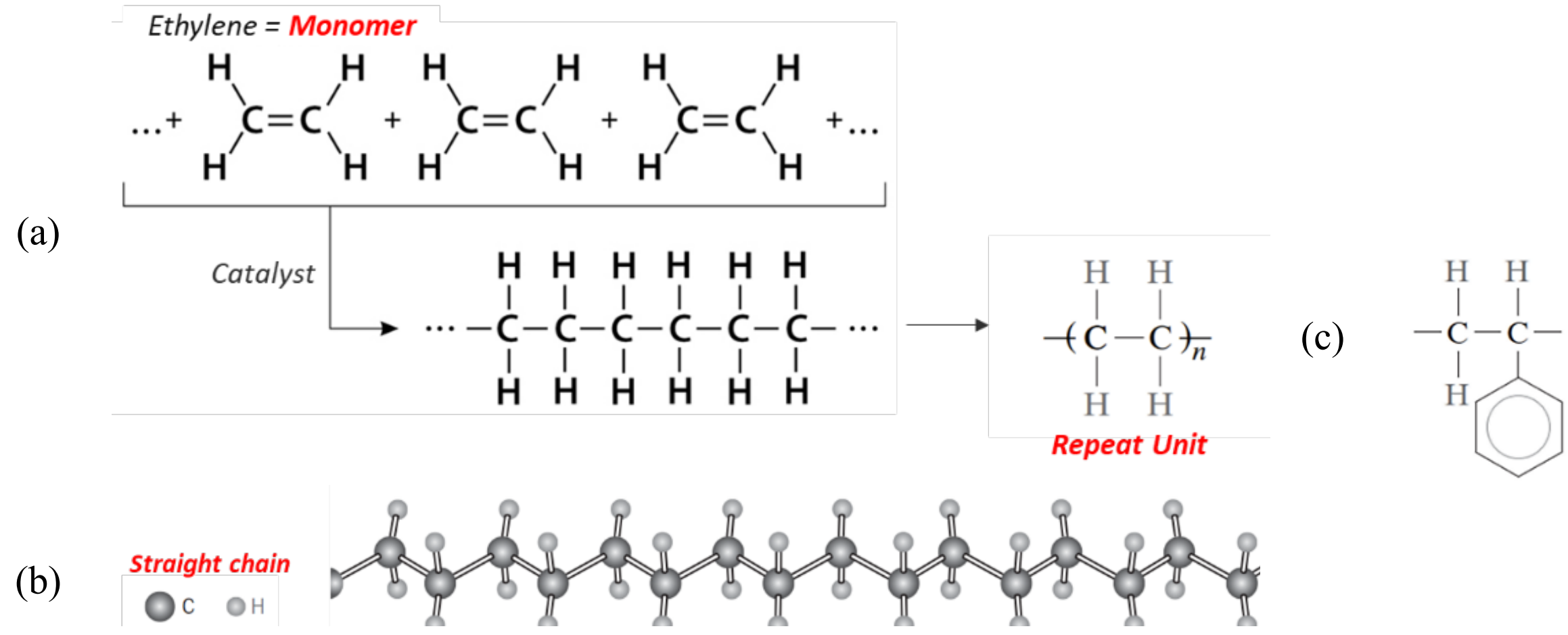

**Fig. 1:** (a) Schematic of PE chain and repeat unit, (b)3D zigzag backbone structure and (c) PS repeat unit with phenyl side group [1].

Polymer synthesis typically follows one of two mechanisms: addition or condensation polymerisation. Addition polymerisation proceeds through initiation, propagation, and termination without releasing by-products. Examples include PE, PS, and acrylics. Condensation polymerisation involves stepwise reactions of functional groups with the release of small molecules (e.g., water or methanol).

Molecular weight ($M_w$) strongly influences state and rigidity (i. e. low values lead to liquids, intermediate values to soft resins, and higher values to rigid solids). Polymerisation produces a distribution of chain lengths, described by the number-average ($\bar{M}_n$) or weight-average ($\bar{M}_w$) molecular weight. Alternatively, the degree of polymerisation (DP) relates $\bar{M}_n$ of the polymer to that of the repeat unit (Fig. 2). A narrow distribution corresponds to more uniform properties in the final material, whereas a broader distribution—though often considered undesirable—can improve processability, as shorter chains act as lubricants that ease the movement and alignment of longer chains.

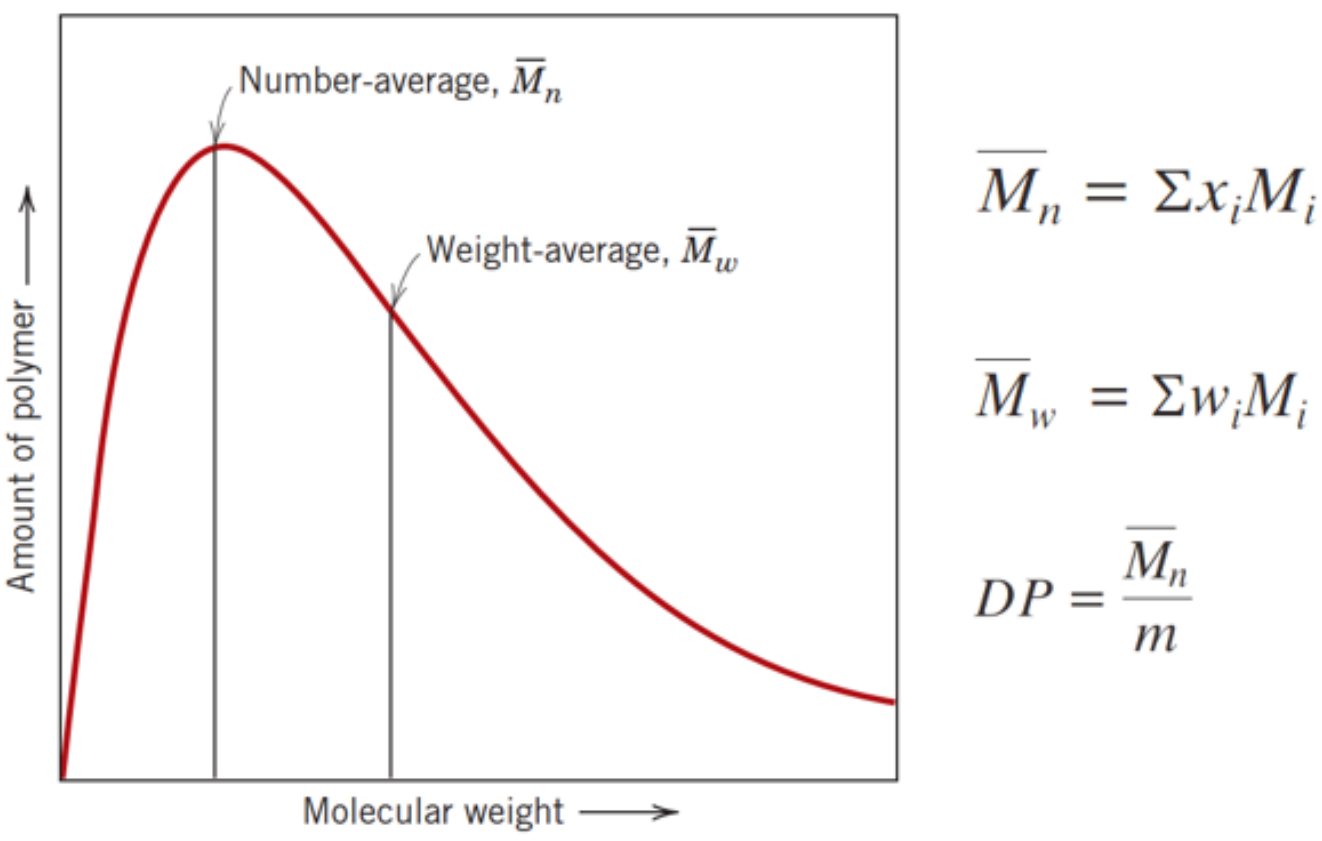

**Fig. 2:** Number- and weight-average molecular weights. "$M_i$" is the molecular weight of size range "$_i$" while "$x_i$" is the mole fraction of chains in that range and "$w_i$" is the corresponding weight fraction [1].

Structurally, polymers may adopt four basic chain architectures: linear (chains free to move relative to one another), branched, cross-linked (chains connected at specific points), and network polymers, formed when cross-linking is extensive and three-dimensional as summarized in Fig. 3. Polymers may also be classified according to their thermal behaviour as either thermoplastics or thermosets. Thermoplastics, such as PE and polypropylene (PP), consist mainly of linear or branched chains with limited cross-linking. They soften when heated and harden upon cooling in a reversible process that can be repeated many times, which makes them flexible and recyclable. In contrast, thermosetting polymers, including epoxies and phenolic resins, contain extensive cross-linking or network structures that form during curing. Once set, they cannot be remelted, and reheating only causes degradation. This irreversible behaviour confers high dimensional stability and resistance to creep, but at the expense of recyclability.

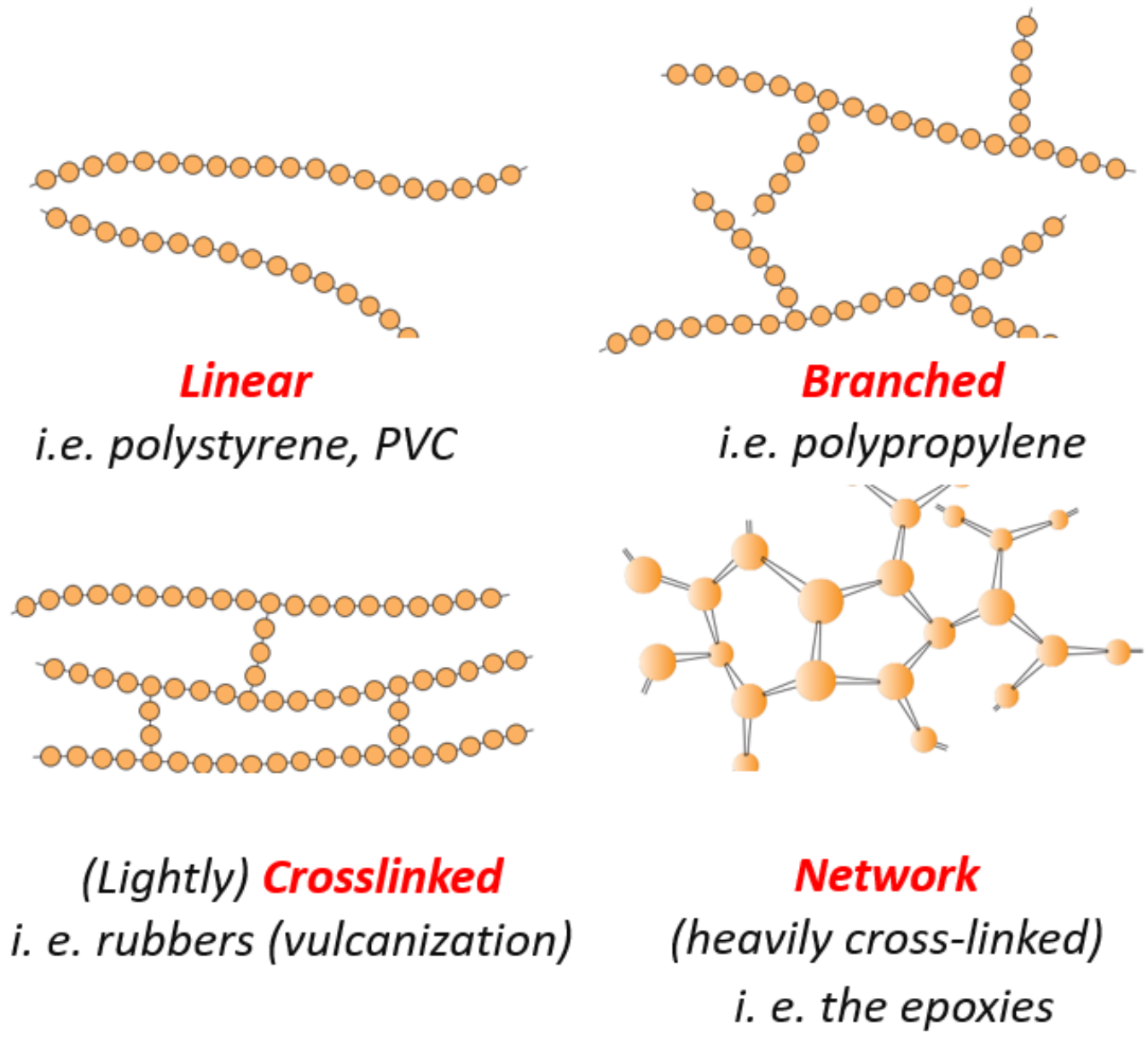

**Fig. 3:** Polymer chain structures: linear, branched, cross-linked and network [1].

The polymers described up to now are generally referred to as homopolymers because they are formed from a single type of monomer, but different monomers can also be combined within the same chain to form copolymers, which can be arranged in various sequences as shown in Fig. 4. In such cases, the degree of polymerisation is defined through the average molecular weight of all repeat units.

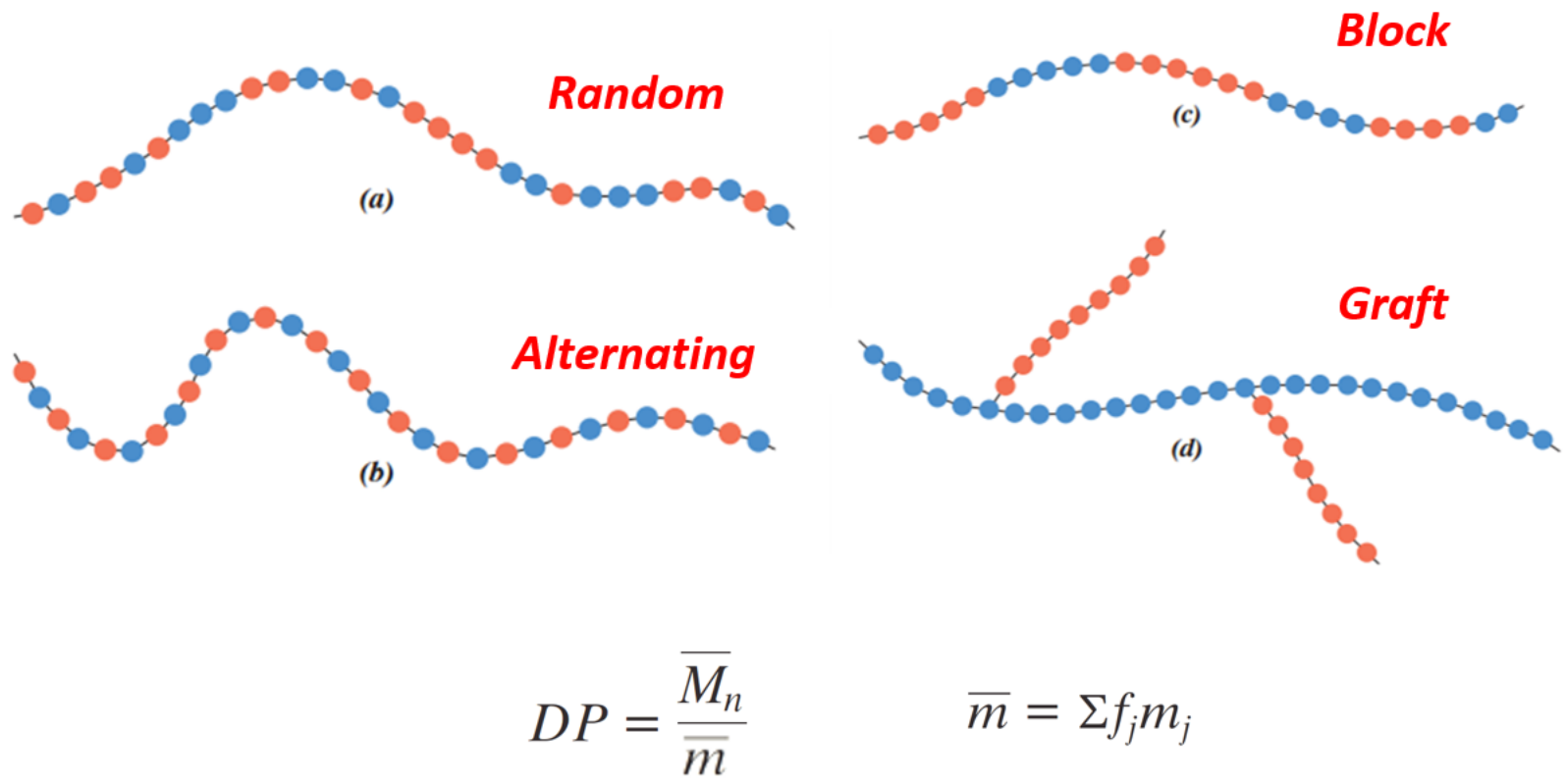

$$DP = \frac{\overline{M}_n}{\overline{m}} \qquad \overline{m} = \Sigma f_j m_j$$

**Fig. 4:** Copolymers. Arrangements of different monomers within a chain and formula for degree of polymerisation where "$f_j$" and "$m_j$" are the mole fraction and molecular weight of repeat unit "$_j$" [1].

### 1.3 Crystallinity in Polymers

Polymers may be fully amorphous but never entirely crystalline. The length and complexity of their chains, together with twisting, coiling, and local disorder, inevitably create amorphous region. Polymers are therefore classified as amorphous and semicrystalline (Fig. 5). Crystallinity is usually quantified from density measurements by comparison with reference amorphous and crystalline states and may reach up to 95–98% under controlled laboratory conditions. Several factors influence the degree of crystallinity: chain length (longer chains promote stronger intermolecular forces), branching (which hinders efficient packing), and interchain interactions (e. g. hydrogen bonding, van der Waals forces, or cross-linking). Increased crystallinity enhances rigidity, tensile strength, and opacity due to light scattering, while amorphous polymers are weaker, more deformable, and often transparent. Thermoplastics may be either amorphous or semicrystalline, whereas thermosets are exclusively amorphous, since extensive cross-linking prevents chains from reorganising into ordered crystalline domains.

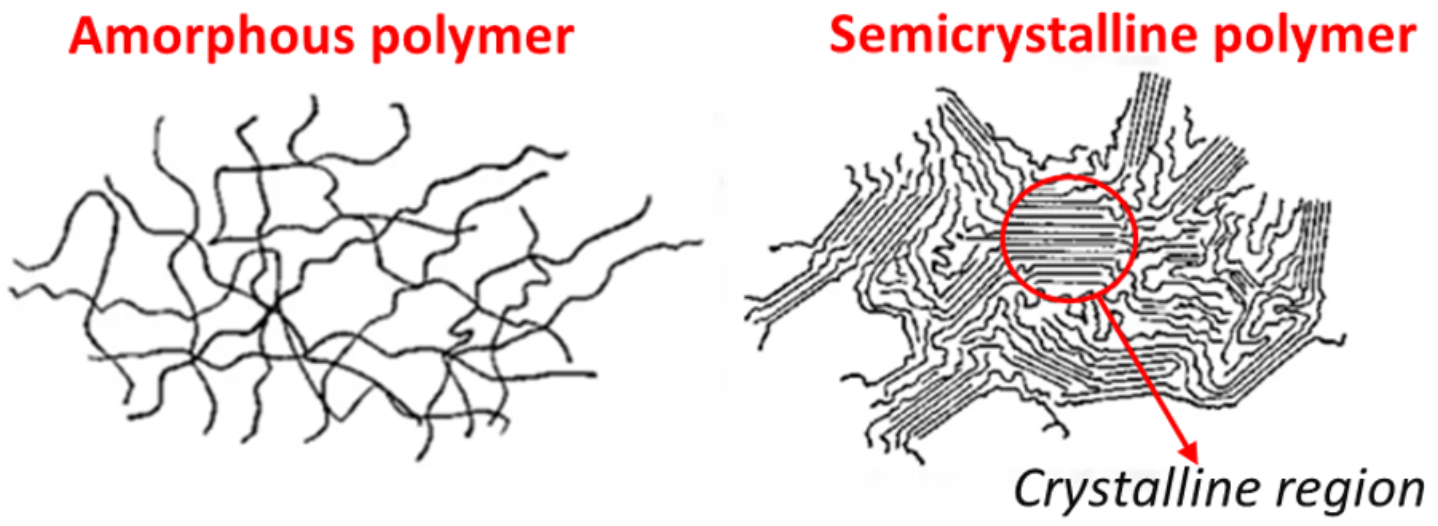

**Fig. 5:** Amorphous and semicrystalline polymers: schematic representations of chain organisation.

At the molecular scale, polymer crystals follow the chain-folded lamellar model, in which long chains align perpendicular to the lamellar surface but fold back to fit within the limited crystal thickness. On a larger scale, lamellae can assemble into spherulites—radially growing, roughly spherical aggregates that form during crystallisation from the melt. Spherulites consist of lamellae arranged around nucleation centres, separated by amorphous regions as represented in Fig. 6(a). They grow outward until impingement with neighbouring spherulites occurs, producing a polycrystalline texture (see Fig. 6(b)). The kinetics of crystallisation from the melt are frequently described by the Avrami equation, which relates the crystallized fraction to time through system-specific constants. The rate decreases with increasing molecular weight but can be accelerated and refined by nucleating agents, which reduce spherulite size. The resulting two-phase morphology plays a decisive role in the physical behaviour of semicrystalline polymers, controlling among others stiffness, strength, permeability, optical transparency, and resistance to radiation.

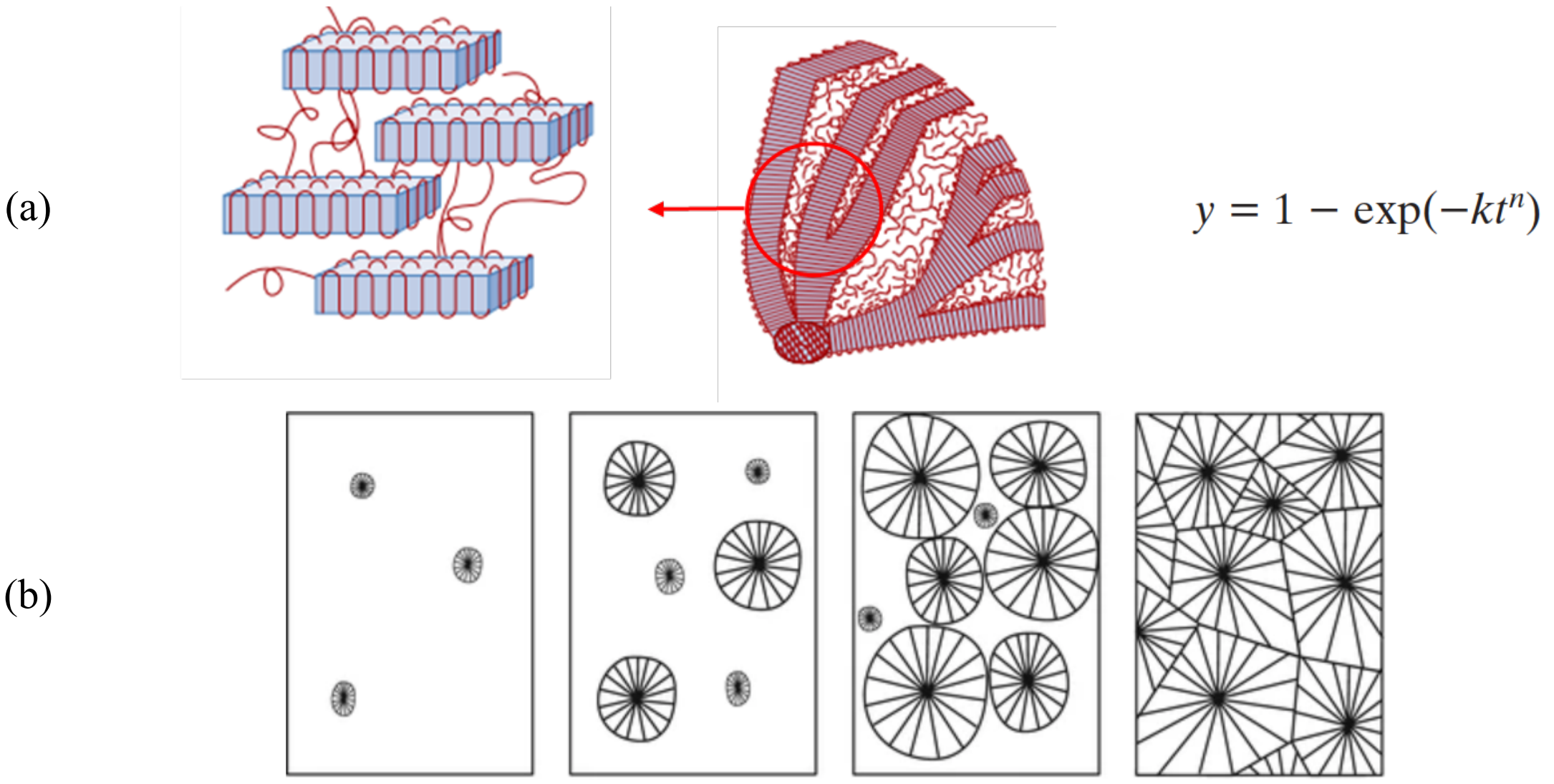

**Fig. 6:** (a) Spherulite structure and Avrami equation where "k" and "n" are time-independent constants specific to the crystallising system [1]; (b) Spherulites growing radially until another spherulite is encountered [3].

### 1.4 Thermal Transitions and Physical Behaviour

Two thermal transitions are of primary importance for polymers: the glass transition temperature ($T_g$) and the melting temperature ($T_m$). Crystallisation occurs between these limits. Unlike metals or ceramics, polymers melt over a temperature range rather than at a sharp point. Melting corresponds to the loss of order as chains fall out of their crystal structures and become disordered in the liquid state. The observed melting behaviour depends on factors such as crystallisation temperature or annealing treatment, lamellar thickness, heating rate, and the presence of impurities or imperfections (thicker lamellae and faster heating raise $T_m$, while defects lower it).

The glass transition is a distinct transformation that affects amorphous polymers or the amorphous fraction of semicrystalline ones. On cooling from the melt, chains lose mobility, and the material becomes rigid (glassy state), while retaining the disordered molecular structure of a liquid. $T_g$ therefore marks the transition between the glassy and rubbery states. Together, $T_g$ and $T_m$ define the lower and upper temperature limits for many applications, particularly in semicrystalline polymers where only the amorphous phase undergoes the glass transition and only the crystalline phase undergoes melting. Some polymers are used in their glassy state (below $T_g$), where they behave as hard and brittle (e.g., PE), or in the rubbery state (above $T_g$), where they are soft and flexible (e.g., polyisoprene). In general, both $T_g$ and $T_m$ increase with chain stiffness: the more rigid the backbone, the higher the transition temperatures.

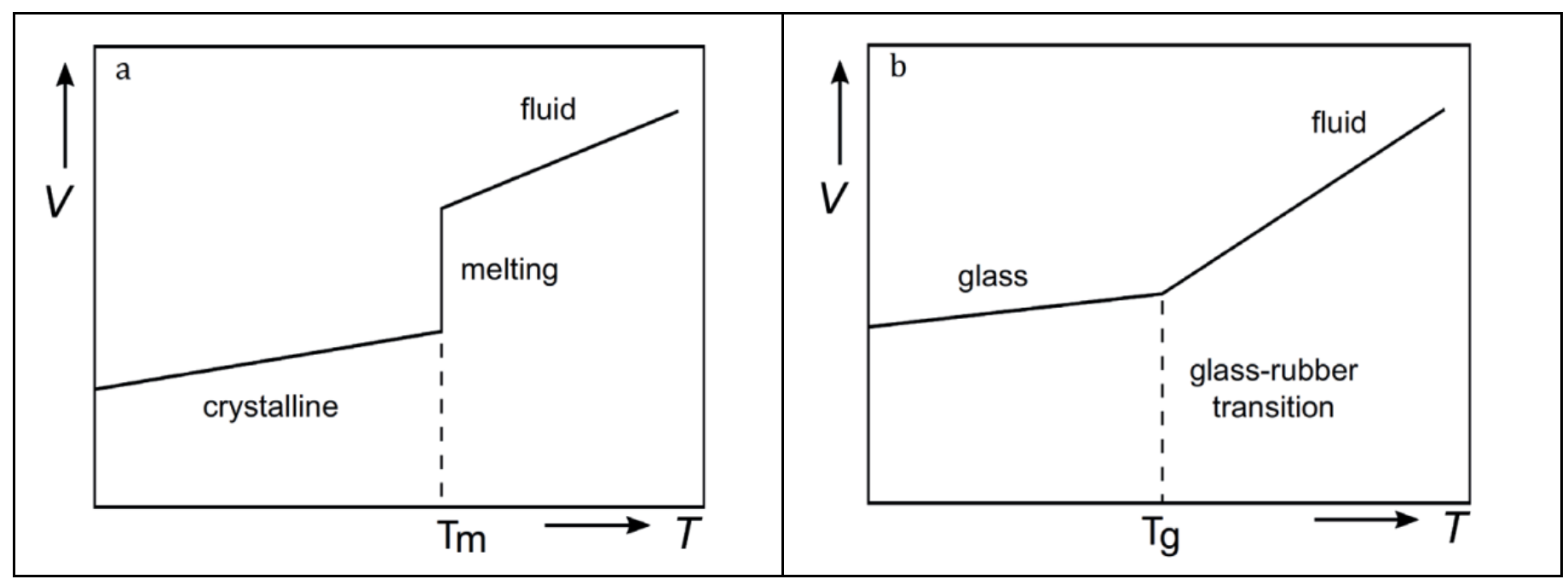

**Fig. 7:** Volume–temperature behaviour: crystalline material (left) and amorphous polymer (right) [4].

Viscosity and viscoelasticity strongly influence both processing and long-term performance. Viscosity decreases sharply with temperature and is critical in applications such as adhesives where sufficiently low viscosity is needed to wet surfaces, spread uniformly, and achieve intimate contact before curing. A case of inadequate viscosity was observed in the ATLAS Inner Tracker (ITk), where silicon modules are bonded to carbon-fibre stave cores using the SE4445 adhesive. In addition to providing mechanical fixation, the adhesive must ensure thermal conductivity to titanium cooling pipes carrying $CO_2$ at -40°C. In some modules, insufficient adhesive spreading led to cracks of the silicon sensors during later manufacturing or testing steps. Optical microscopy and computed tomography (CT) evaluation (Fig. 9) confirmed uneven adhesive layers and voids that compromised both the bond and the thermal path and ultimately reducing component reliability.

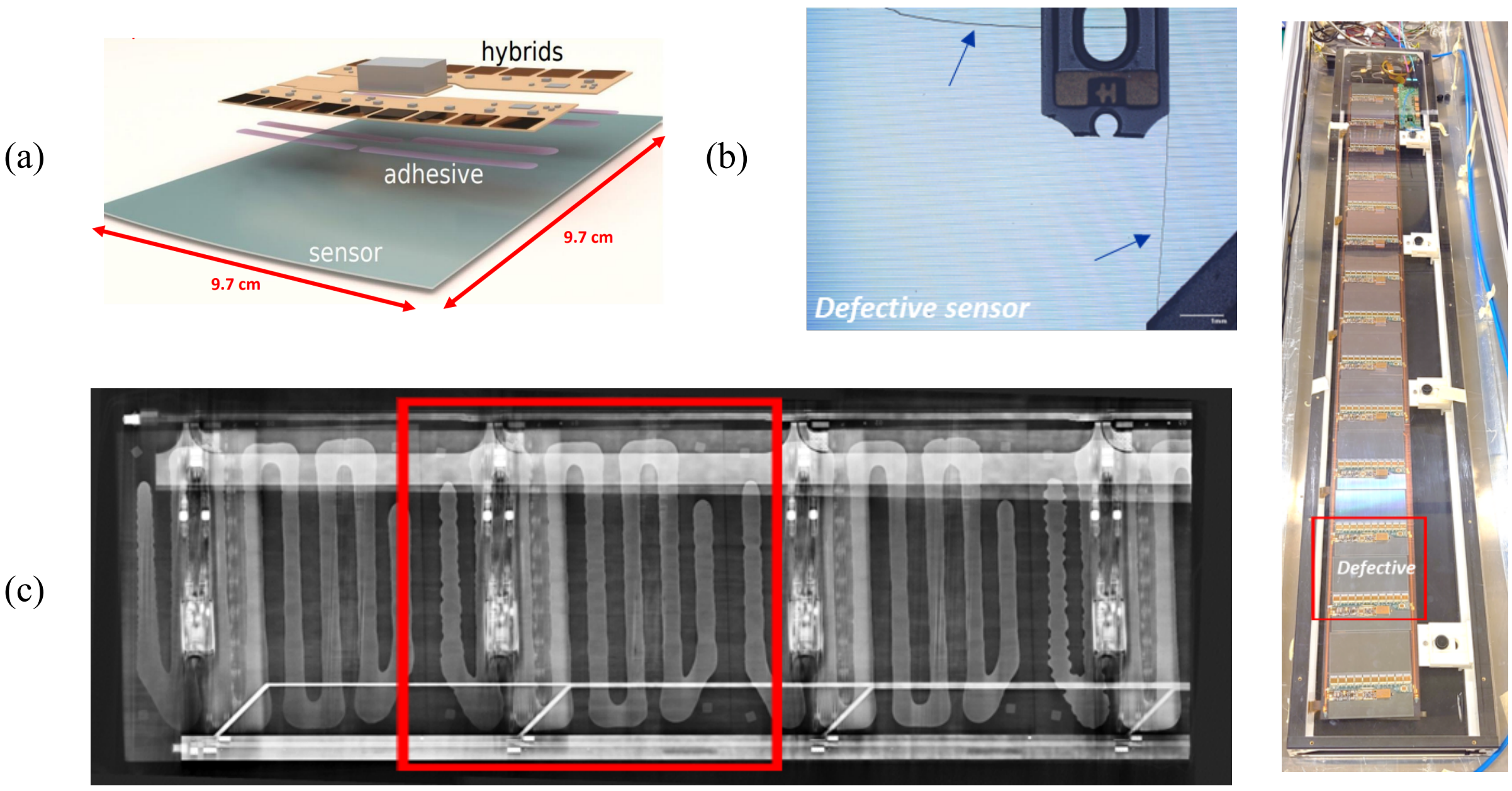

**Fig. 8:** Adhesive application in ATLAS ITk modules. (a) Concept of a strip module (adapted from [5]); (b) Microscopy image of a defective silicon sensor showing cracks after thermal testing; (c) CT images of glue distribution revealing insufficient spreading at the interface [6].

Viscoelastic behaviour—intermediate between ideal elastic solids and viscous fluids—causes creep and stress relaxation under load. Creep, the gradual time-dependent deformation, is particularly critical in structural and adhesive applications, especially under thermal cycling or irradiation.
Another key feature of polymers is their high, nonlinear coefficient of thermal expansion (CTE), which changes markedly near $T_g$. Mismatches in CTE between polymers and metals can induce residual stresses in composites and bonded joints. This is a crucial design factor for accelerator components that combine polymers or polymer–matrix composites with other materials.

### 1.5 Mechanical Behaviour of Polymers

The mechanical properties of polymers are characterised by the same parameters as metals, including modulus of elasticity, yield strength, and tensile strength. However, polymer response depends strongly on strain rate, temperature, and environment (e.g. humidity, oxygen, or solvents). Standard stress–strain curves reveal three typical behaviours: brittle (little plasticity before fracture), plastic (metal-like, with elastic, yield, and strain-hardening regimes), and elastomeric (large recoverable strains at low stresses). Elastic response arises from intermolecular forces and chain entanglement or cross-link networks.

Modulus can be further increased by fillers (e.g. glass fibres), greater cross-link density (vulcanisation), or molecular orientation (e.g. high-performance polyethylene fibres with moduli approaching 200 GPa). Increasing temperature generally reduces modulus and tensile strength while enhancing ductility, with pronounced effects near the glass transition. For example, Plexiglas is brittle at 4°C but plastically deformable at 50–60°C.

This temperature dependence is critical in accelerator technology. Epoxy resins are widely used to impregnate superconducting magnet coils, where in addition to providing electrical insulation they confer mechanical stability, stiffness, and protection during handling and operation.

Long-term reliability requires low viscosity for void-free impregnation of long coils with complex geometry, and good adhesion to minimise bond failure with other components. A moderate pot life is also necessary to allow sufficient processing time before viscosity rises, while low-temperature curing helps reduce internal thermal stresses (see Fig. 9a). In service, toughness and resistance to cracking are essential at both, room temperature and cryogenic conditions (1.9 K), with values varying significantly depending on the resin system, as shown in Fig. 9b [7–8]. Finally, resins must retain acceptable strength and flexibility after irradiation, since they are typically the first magnet material to incur radiation damage.

Continuous improvements and alternative formulations have enhanced resistance to thermal shock and cracking, though often at the expense of other parameters such as viscosity, adhesion, or curing behaviour [9–10].

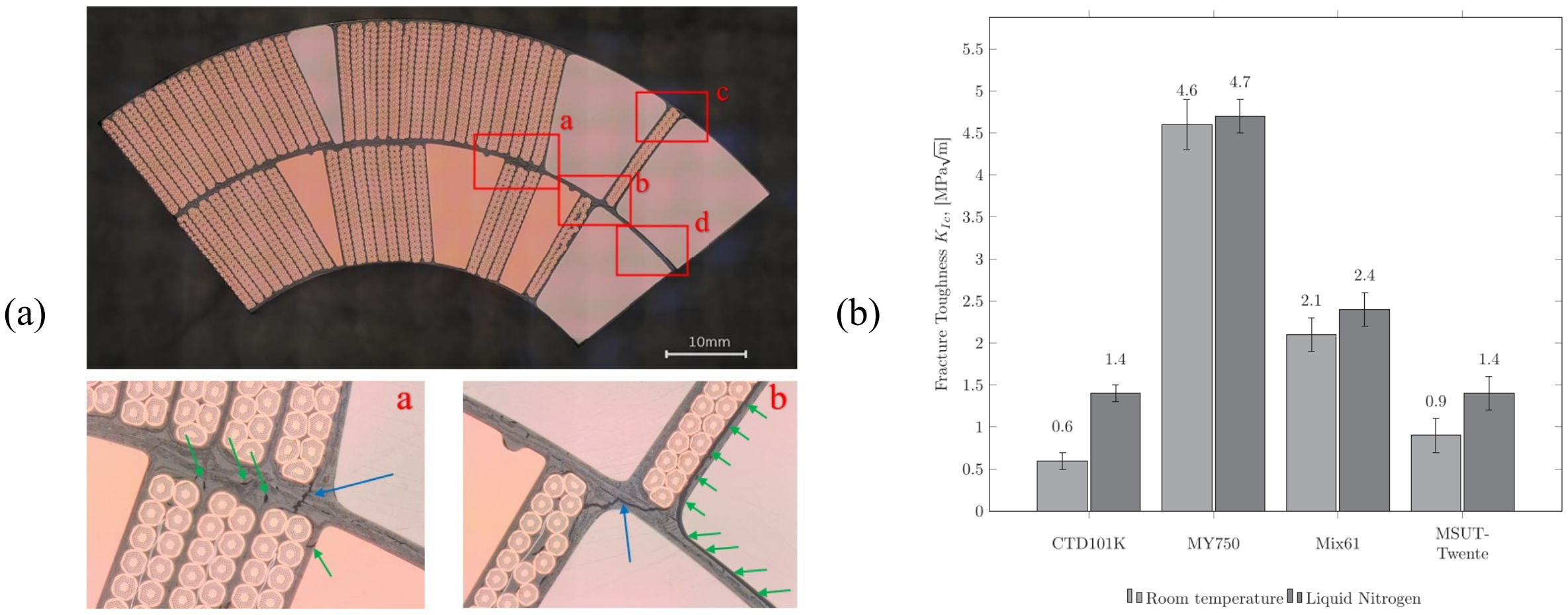

**Fig. 9:** (a) Coil section from an 11 T dipole magnet impregnated with CTD101K epoxy, showing defects such as shrinkage cavities, interlayer cracks, and decohesion [8]; (b) Fracture toughness of various epoxy systems measured at room temperature and liquid-nitrogen temperature [10].

### 1.6 Additives

To adjust the properties discussed previously beyond what can be achieved through structural modification alone, additives are commonly incorporated into polymer formulations to enhance serviceability. Typical examples include fillers to increase stiffness and wear resistance, plasticizers to improve flexibility, stabilizers to counteract thermal or UV degradation, colourants for identification, and flame retardants to improve fire resistance, thereby enabling the safe use of polymers in demanding environments such as underground facilities.

### 1.7 Radiation on Polymers

The behaviour of polymers under ionising radiation is a critical factor in their use for accelerator components operating in high-radiation areas. Radiation alters polymer structure causing gas production, acid formation, oxidation, and separation of multi-phase materials like greases making and making materials softer or harder through processes such as chain scission and cross-linking. These molecular changes appear macroscopically as embrittlement, reduced elongation, colour change and altered thermal or dielectric properties (see Fig. 10(a)). While these changes are often correlated with the total absorbed dose, the outcome also depends strongly on irradiation conditions such as radiation type, dose rate, temperature, humidity, oxygen content, and simultaneous mechanical or electrical stresses. These factors may act synergistically causing more severe degradation than the sum of individual effects.

Understanding the radiation tolerance of lubricants, adhesives, elastomeric sealants, cable insulations, and vacuum components is essential to minimise the risk of failure and to reduce design safety margins. At CERN, irradiation tests are systematically performed on both commercial plastics and tailor-made components. Extensive studies carried out from the 1960s to the 2000s were compiled in the well-known Yellow Reports [11-13]. Today, dedicated projects such as R2E and CARE [14–15] organise up to ten irradiation campaigns per year in external facilities or in-house installations, where polymers are typically exposed to total doses in the 0.1–20 MGy range at dose rates of several kGy/h. Radiation effects generally become measurable above about 0.1 MGy (Fig. 10(b)). Exposures exceeding 1 MGy require materials with experimentally verified resistance, and for doses above 10 MGy the use of organic materials is generally discouraged unless complemented by frequent monitoring and predefined failure-management strategies.

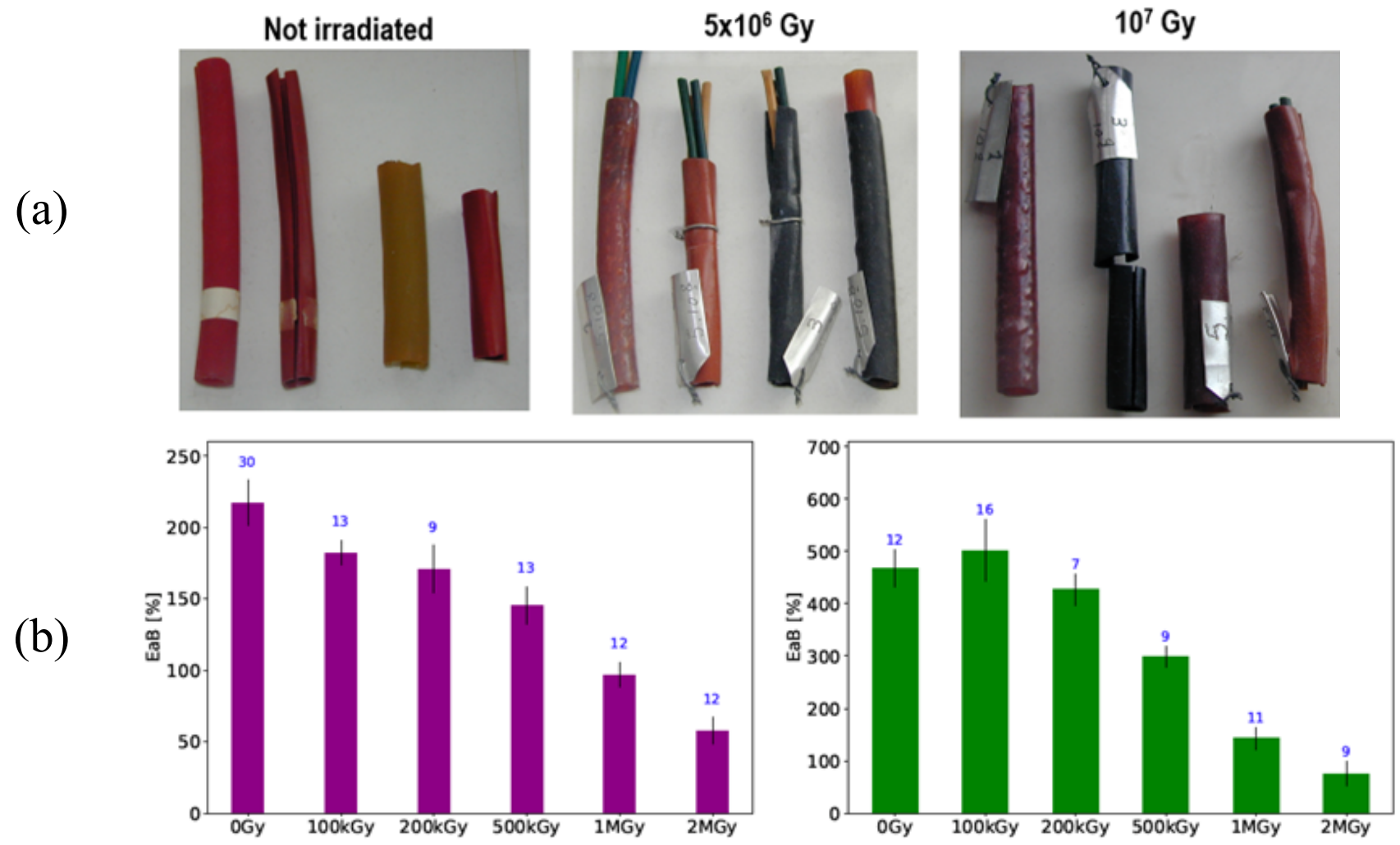

**Fig. 10:** (a) Representative examples of installed cables subjected to different total absorbed dose and (b) Elongation at break (EaB%) vs absorbed dose on cable jacket (EVA) in purple and insulation (XLPE) in green.

A special case is that of superconducting magnets, which represent one of the most demanding applications for polymers in accelerator facilities. In these systems, radiation effects are compounded by operation at cryogenic temperatures, creating a unique combination of stressors. Understanding the ageing of impregnation resins under relevant conditions—including irradiation source, temperature, oxygen content, and formulation additives—is therefore pivotal to ensure reliable magnet design and long-term performance [16].

## 2 Composite Materials

### 2.1 Introduction

Composite materials are a relatively recent family of engineered materials that have become essential wherever high stiffness-to-weight ratios and tailored performance are required. Their emergence in the mid-20th century, marked by the production of glass-fibre reinforced polymers, represented a turning point in materials science. Since then, composites have found widespread use in aerospace, energy, civil engineering, and transport, enabling lighter, stronger, and more versatile structures than conventional metals or ceramics [17].

## 2.2 Definition and classification

A composite is any material made from two or more distinct phases, separated by a well-defined interface and chemically dissimilar from each other. Typically, a continuous matrix binds together a dispersed phase such as fibres, particles, or laminates. The overall properties depend not only on the intrinsic characteristics of these constituents, but also on their relative amounts, the quality of the interface, and the geometry, size, distribution, and orientation of the reinforcement.

Composites exist in nature—for example, wood and bone—or in multiphase polymers and alloys such as duplex steels. In the present context, however, the term refers to artificially engineered multiphase materials, designed to achieve property combinations unattainable with metals, ceramics, or polymers alone. Classification is generally based on the matrix type (polymeric, metallic, or ceramic), the reinforcement (fibres, particles, or structural laminates), or the structural arrangement. The superior performance of carbon-fibre reinforced polymers (CFRPs) and glass-fibre reinforced polymers (GFRPs) is compared with conventional materials in Fig. 11 where the composites occupy a region of the charts that demonstrates their outstanding strength-to-weight ratios and balanced toughness, explaining their widespread adoption in high-performance applications [18]. More recently, nanocomposites, where nanoscale phases are dispersed within a matrix, have emerged as a rapidly growing class with unique mechanical, electrical, and functional properties.

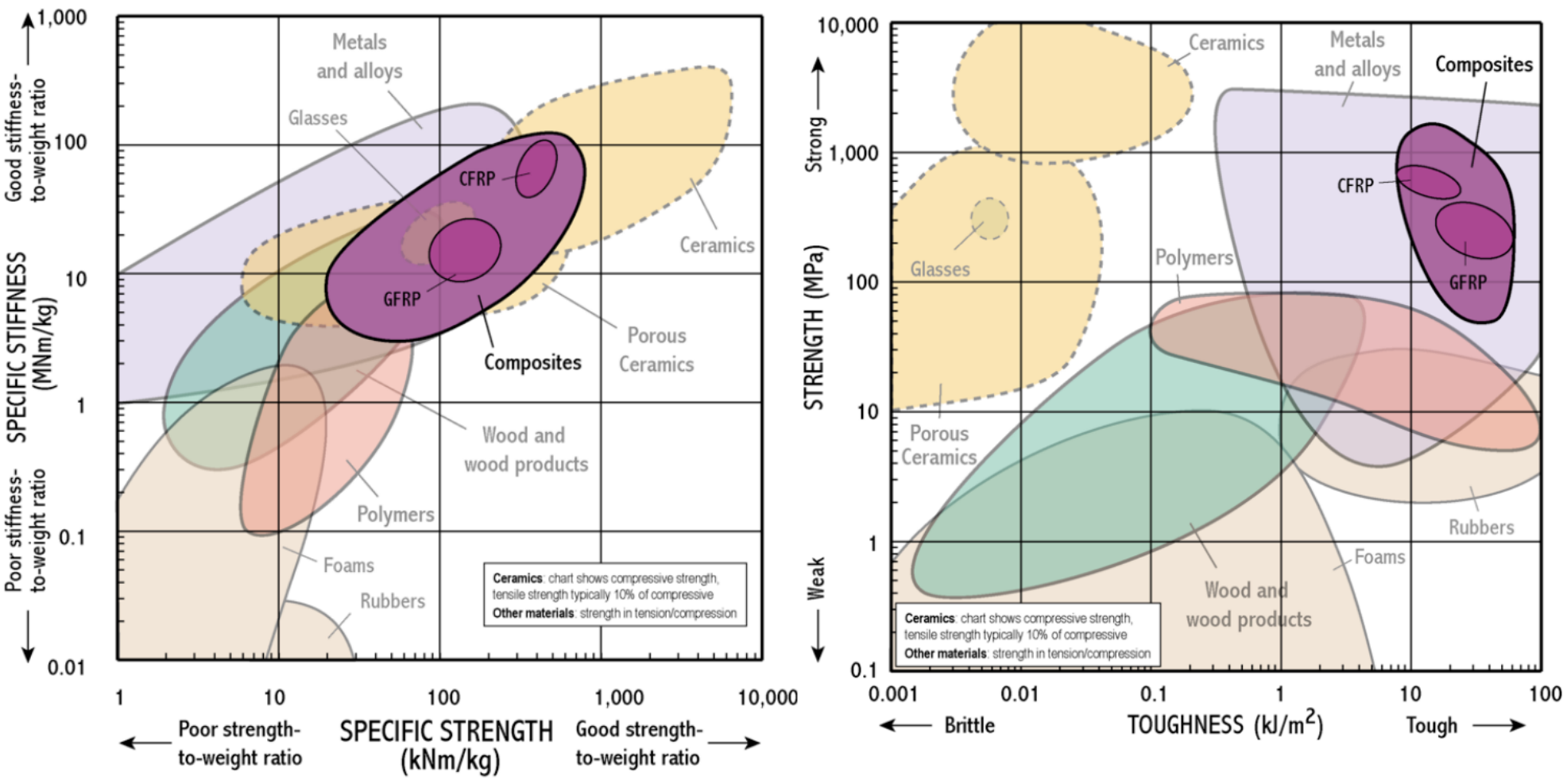

**Fig. 11:** Property maps comparing composites with conventional materials: (left) specific stiffness versus specific strength, and (right) strength versus fracture toughness.

While generally they are classified by matrix type (polymeric, metallic, or ceramic), by reinforcement (fibres, particles, or structural laminates), or by structural arrangement, not all categories are equally relevant to accelerator technology. In the present work, emphasis is placed on those families most important for high-energy physics applications.

## 2.3 Fibre-Reinforced Composites (FRCs)

Fibre-reinforced composites derive their exceptional mechanical performance from the properties of the fibres and the efficiency with which the surrounding matrix transfers load to them. The fibre type, orientation, and volume fraction are decisive parameters, as proven in Table 1. For effective strengthening, fibres must exceed a critical length—on the order of 1 mm for glass and carbon fibres, corresponding to ~20–150 times their diameter—so that stresses are efficiently transmitted through the matrix–fibre interface. Short fibres below this length provide only limited reinforcement, while continuous fibres deliver the highest performance.

**Table 1**: Effect of reinforcement volume fraction on selected composite properties [1]

| Property | Unreinforced | Value for given amount of reinforcement (vol.%) | | |
|---|---|---|---|---|
| | | 20 | 30 | 40 |
| Tensile strength (MPa) | 59–62 | 110 | 131 | 159 |
| Modulus of elasticity (GPa) | 2.24–2.345 | 5.93 | 8.62 | 11.6 |
| Elongation (%) | 90–115 | 4–6 | 3–5 | 3–5 |

As shown in Fig. 12, unidirectional plies, provide maximum stiffness and strength along the fibre axis, while properties transverse to the fibres are governed by the matrix. Depending on the application 1D arrangement can be extremely useful if the load in-service is oriented along the fibre's axis while if the component needs strength and stiffness in several direction (is the case in some pressure vessels) the solution is to stack multiple layers with different fibre orientations 2D (laminates, fabrics) or 3D (braids, stitched or woven structures) [19–20]. The resulting stack is called laminate and can have quasi-isotropic properties if enough layers with the correct orientation are stacked. Post-lay-up processing techniques such as autoclave moulding, pressure-bag moulding, and vacuum-bag moulding are employed to reduce porosity and improve laminate quality.

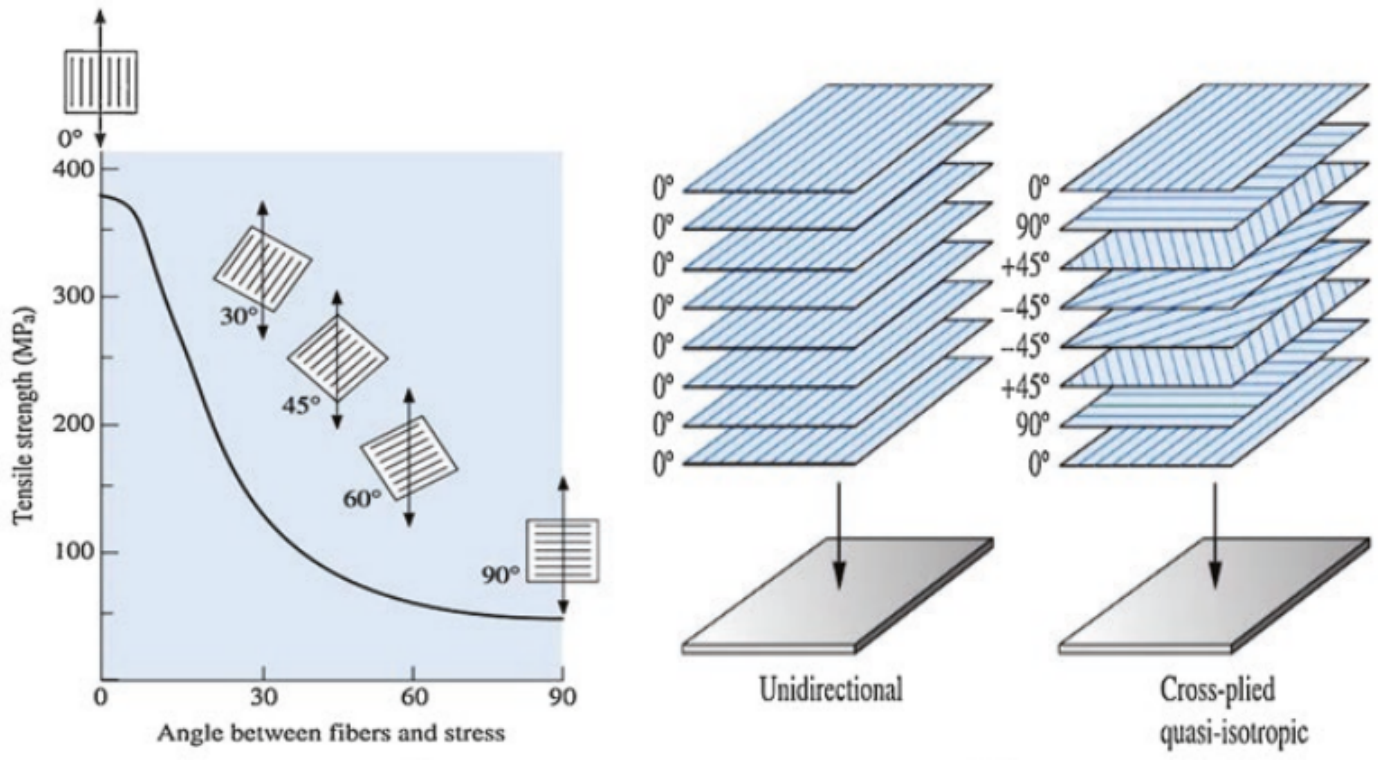

**Fig. 12:** Effect of fibre orientation on the tensile strength of E-glass/epoxy composites, comparing unidirectional and cross-plied quasi-isotropic laminates [19].

This approach is for example applied at CERN to produce support structures for the ATLAS Inner Tracker (ITk), whose goal is to record particle trajectories with extreme precision while minimising material that could disturb them. CFRCs are employed to manufacture exceptionally thin, robust, and dimensionally accurate rings, as illustrated in Fig. 13. Precise knowledge of the experimental mechanical properties of the material—both in the “pre-preg” laminate format and after manufacturing—is essential. Simulation and design studies, including the development of all necessary tooling and moulds, are followed by prototype production and testing at the CERN Composites Laboratory to validate the manufacturing route and to ensure that service stresses align well with the directions of highest strength.

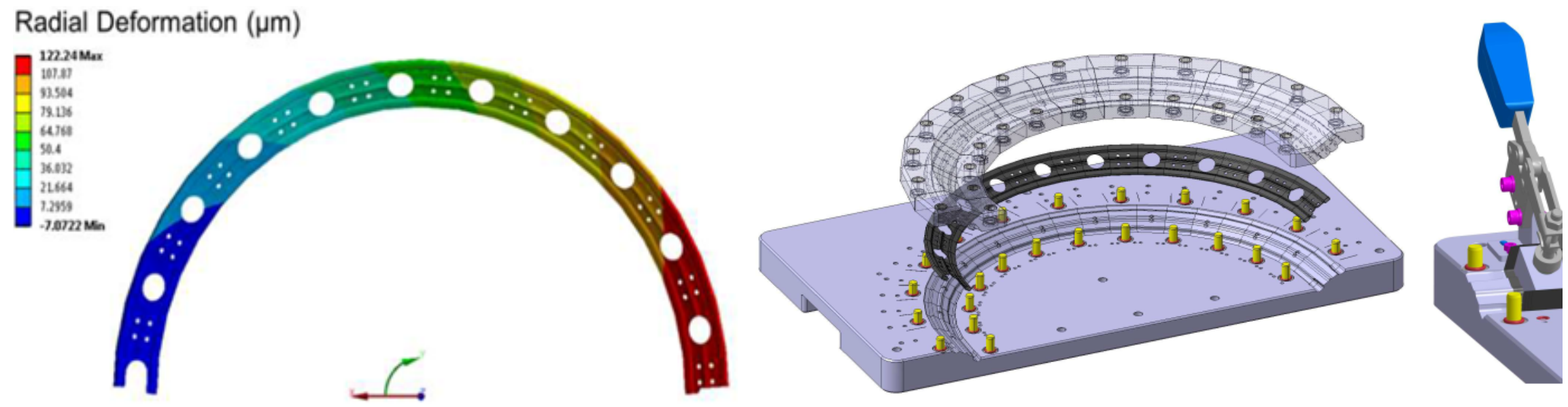

**Fig. 13:** Conceptual design of ATLAS Inner Tracker structural rings in CFRP composite and associated tooling developed at the CERN Composites Laboratory.

The mechanical characteristics of a fibre-reinforced material depend ultimately on the quality of the fibre–matrix interface. The interfacial area in composites is extremely large (on the order of 100,000 $m^2/m^3$) and becomes even more significant as fibre diameter decreases, as in nanoparticle-reinforced systems. The quality is influenced by factors such as wettability (the ability of the matrix to spread on the fibre surface), mechanical interlocking (arising from surface roughness), and residual stresses generated during processing (e.g. due to differences in thermal expansion coefficients, cooling shrinkage, or water absorption) [21].

The importance of the fibre–matrix interface is clearly illustrated in a CERN case study where inadequate interfacial quality led to defective components. Cooling adaptors machined from aluminium–carbon fibre ($Al/C_f$) metal–matrix composites, containing ~50 vol% chopped fibres randomly oriented in the XY plane, were tested for use in the outer tracker of the CMS detector, which operates silicon sensors at -30°C. The material was selected to match the thermal expansion coefficient of silicon, ensuring dimensional stability within a few micrometres during thermal cycling, while also being non-magnetic. Two commercial grades were compared: provider A (cast $Al/C_f$) and provider B (sintered $Al/C_f$). Although the block format from provider B was more suitable for machining, significant dimensional variations developed over time [22]. Detailed scanning electron microscope (SEM) investigation revealed galvanic corrosion at the fibre–matrix interface as the degradation mechanism, arising from electrochemical interactions between the carbon fibres and aluminium matrix. The more open microstructure of the powder-metallurgy route appears to have exacerbated this effect, highlighting how processing can influence interfacial integrity and long-term stability [23].

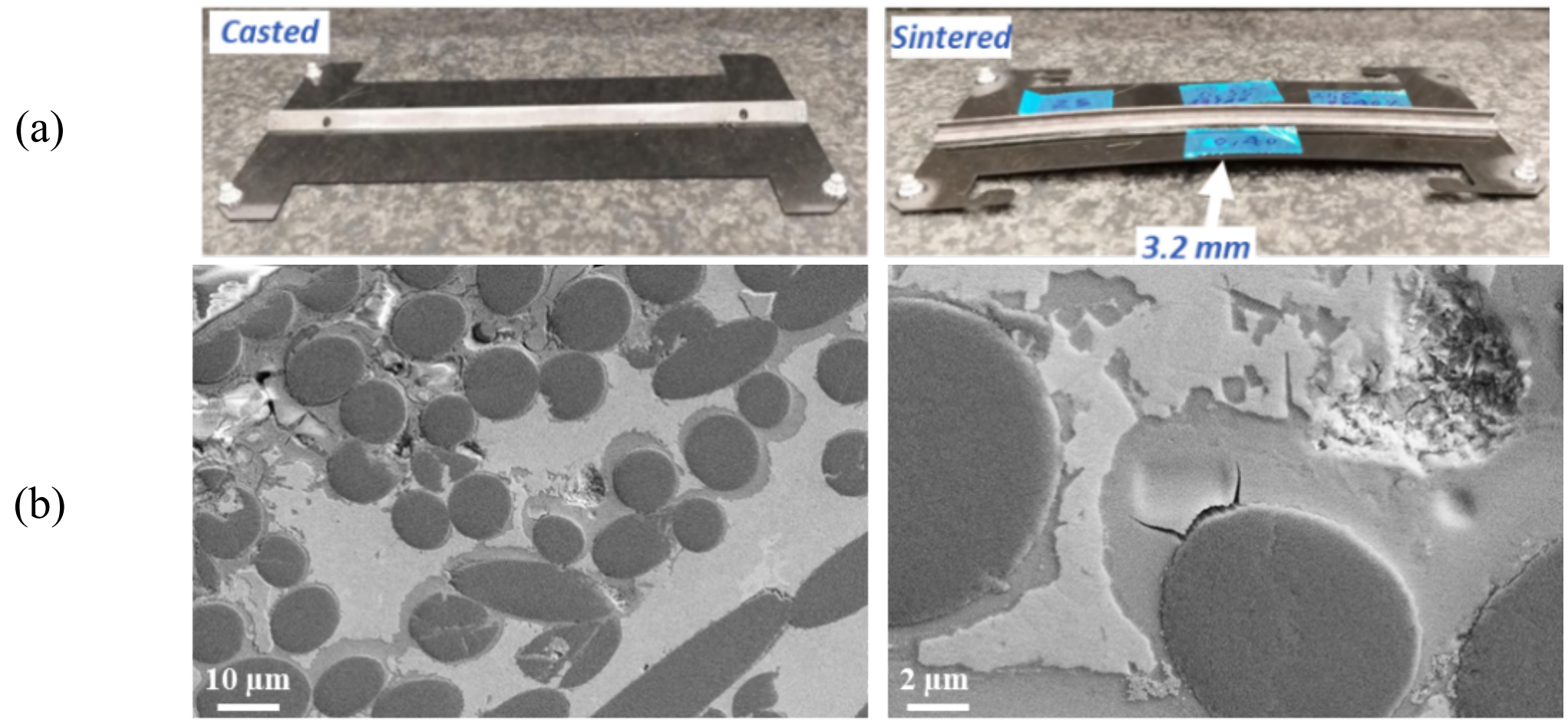

**Fig. 14:** Cooling adaptor for the CMS Outer Tracker made from Al–CF metal–matrix composites (a) cast and sintered variants comparison and (b) SEM images after wet testing revealed signs of galvanic corrosion at the Al–CF interface in sintered samples.

Polymer-matrix composites (PMCs) are the most widely used class of fibre-reinforced materials. The matrix determines the maximum service temperature, since it typically softens, melts, or degrades well below the limits of the reinforcements. Two main polymer families are employed: In one side, thermosetting resins—notably epoxies, polyesters, and polyimides—are common due to their low viscosity before curing, which facilitates impregnation of fibre bundles and complex shapes. Their main drawback is brittleness, especially at cryogenic temperatures. On the other hand, thermoplastics such as PEEK, PPS, and PEI offer greater toughness and a broader operating temperature range, but their higher viscosity complicates processing. Other matrix systems include metal-matrix composites (MMCs), valued for high-temperature capability, non-flammability, and chemical resistance, and ceramic-matrix composites (CMCs), which provide exceptional thermal stability and wear resistance. Both, however, are limited by high cost and processing complexity.

### 2.4 Particle-Reinforced Composites

Particle-reinforced composites can be divided into two classes. Large-particle composites include typical examples such as polymers filled with carbon black particles, such as synthetic rubbers used in tires, or cement-based concretes used as structural materials. In contrast, dispersion-strengthened composites contain very fine particles, typically 10–100 nm in diameter, where strengthening occurs at the atomic or molecular scale. The matrix carries most of the applied load, while the dispersed particles hinder dislocation motion, thereby increasing strength and creep resistance.

At CERN, copper-based materials with dispersed phases are of particular interest. Cu–diamond (CuCD) is a true particle-reinforced composite, exploiting the high thermal conductivity of diamond particles dispersed in a copper matrix to enhance heat dissipation in beam-intercepting devices [24]. In contrast, dispersion-strengthened materials such as Glidcop®, an oxide dispersion strengthened (ODS) copper alloy containing $Al_2O_3$ nanoparticles, are also widely used due to their combination of high electrical and thermal conductivity with improved mechanical strength and radiation tolerance [25–26].

### 2.5 Structural Composites

Structural composites are multilayered and typically low-density systems designed for high stiffness, strength, and stability. Their properties depend not only on the constituent materials but also on the geometry of the structural elements. Sandwich panels, for example, consist of two strong outer skins bonded to a lightweight core (e.g. foams, honeycombs, or balsa wood), providing excellent stiffness-to-weight ratios and high compression resistance. Similarly, laminar composites are built by stacking and bonding fibre-reinforced plies, where the ply orientation can be tailored to achieve quasi-isotropic performance.

A CERN example of a structural composites is Gas Electron Multipliers (GEMs), thin laminates consisting of a polymer foil (typically 50 µm Kapton) metal-coated on both sides with copper (~5 µm) and perforated with a high density of microscopic holes as displayed in Fig. 15. GEMs are CERN patented technology used as stand-alone detectors or as preamplifiers in multi-stage structures. Their composite architecture provides a combination of mechanical robustness, flexibility, and electrical functionality at low cost. They can operate with large overall gains in harsh radiation environments, and their modular construction allows straightforward assembly from prefabricated components. GEMs also offer configurable design options, enabling the optimisation of detector performance for specific experimental requirements.

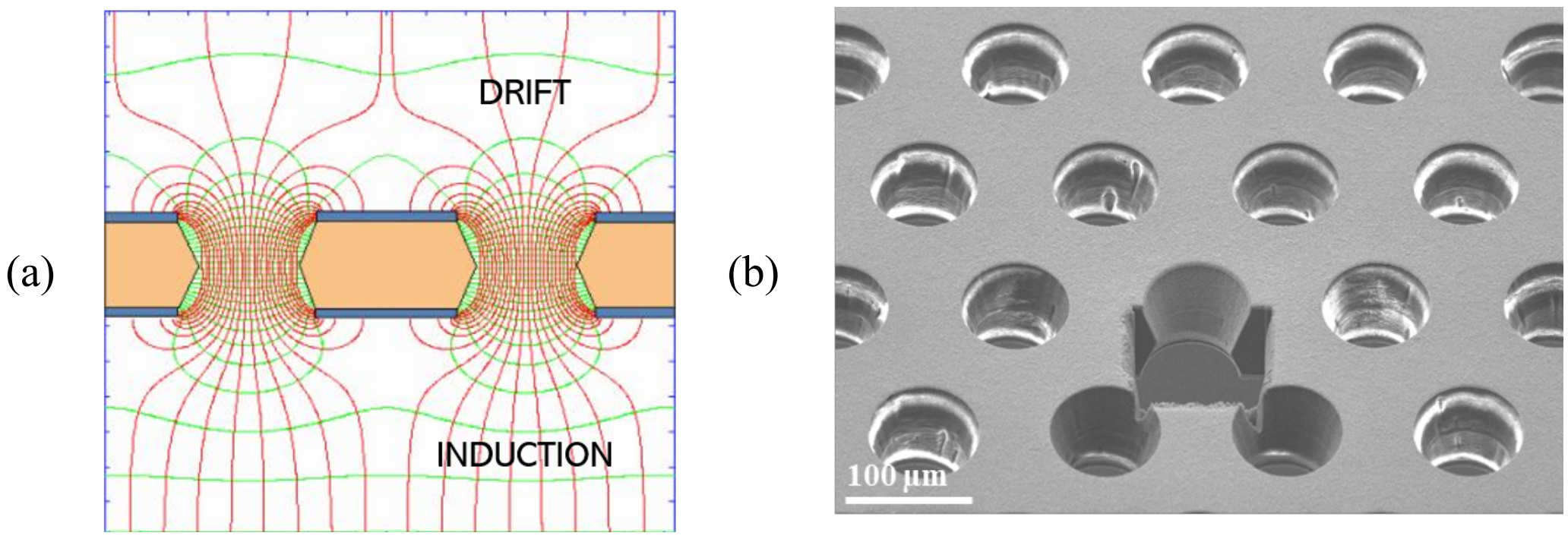

**Fig. 15:** (a) Electric field in the region of the holes of a GEM electrode [27] and (b) focused ion beam (FIB)-SEM cross sectional view on GEM hole as part of provider qualification at CERN.

### 2.6 Nanocomposites

Nanocomposites are a rapidly developing class of materials in which at least one of the constituent phases has dimensions below 100 nm. At this scale, the large surface-to-volume ratio of the dispersed phase means that properties are strongly influenced by the interface quality. Depending on the type and content of the nanophase, these materials can display remarkable improvements in strength, toughness, electrical conductivity, thermal stability, or functional behaviour. Nanoparticle loadings vary widely with application. For example, addition of carbon nanotube concentrations of around 5 wt% can substantially increase strength and stiffness, whereas 15–20 wt% are typically required to reach electrical conductivities sufficient to protect nanocomposite structures from electrostatic discharges. The balance between mechanical reinforcement and functional performance makes careful optimisation of nanoparticle content essential.

Although not a composite material in the classical sense, $Nb_3Sn$ wires with artificial pinning centres (APCs) exhibit a composite-like nanoscale architecture, with a superconducting matrix containing finely dispersed oxide nanoprecipitates, and represent an example of advanced microstructural engineering at CERN, illustrating how nanoscale features can be exploited to tailor functional properties. These materials are being developed for next-generation accelerator magnets, where enhanced current densities are essential to reach the magnetic fields required for high-energy operation. In standard $Nb_3Sn$, grain boundaries act as the primary pinning centres; consequently, materials with finer grains achieve higher current densities. Typical $Nb_3Sn$ grain sizes are ~100–150 nm. Introducing oxygen uniformly during heat treatment refines grain size, enhancing vortex pinning and raising the layer critical current density (Jc) at 4.2 K. Oxygen is supplied from carefully placed internal sources, from which it diffuses effectively during the post-deformation heat treatment [28–30], and the presence of oxides can be evaluated via high-resolution electron microscopy, as illustrated in Fig. 16.

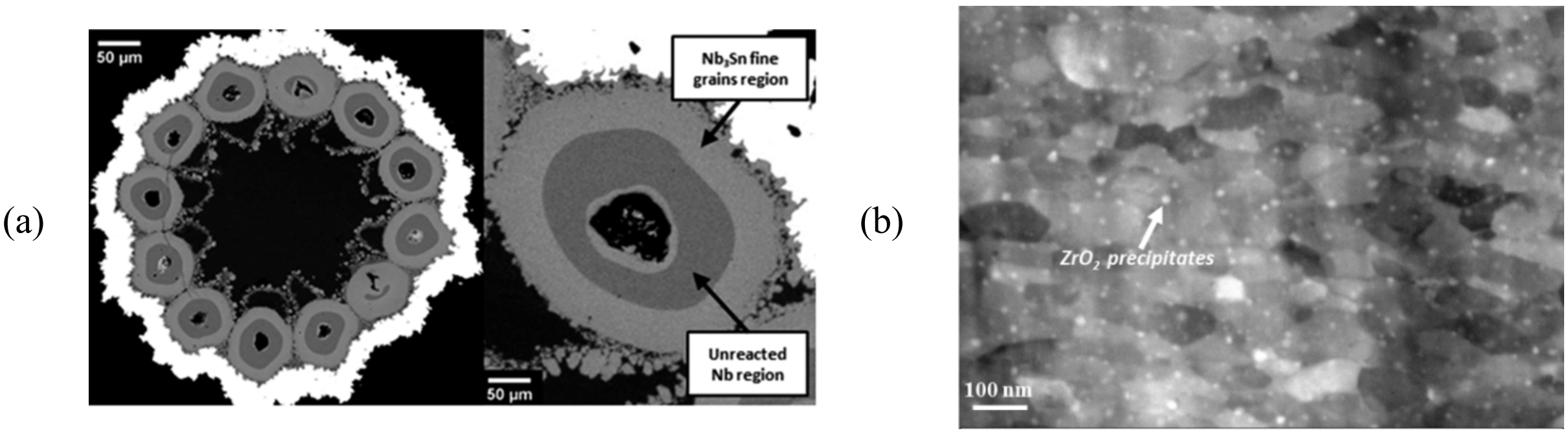

**Fig. 16:** (a) SEM image of a Zr-core oxygen-source $Nb_3Sn$ wire showing unreacted Nb, fine-grained $Nb_3Sn$, and coarse-grained outer regions [30] and (b) scanning transmission electron microscope (STEM) image of APC $Nb_3Sn$ wire highlighting dispersed $ZrO_2$ nanoprecipitates in white.

## 3 Conclusions

This paper provides a general introduction to polymers and composite materials, with emphasis on their application in accelerator and detector technologies. Understanding the structure, intrinsic properties and behaviour of these materials is fundamental to identify optimal solutions for demanding environments like those at CERN that often combine cryogenic conditions, high mechanical stresses, and intense radiation fields. An overview of their characteristics and classification has been given, highlighting the strengths and limitations of each type, key aspects of material selection and design, and in-service considerations. Specific CERN case studies—including adhesives, epoxy resins, fibre-reinforced composites, and nanostructured superconducting wires—have illustrated the practical use and critical importance of these materials for reliable operation in present and future accelerator facilities.

## Acknowledgements

I wish to thank the organisers for the invitation to contribute and the colleagues from the Polymer Laboratory and the Composites Laboratory at CERN and to G. Arnau and S. Sgobba (EN-MME) for the instructive discussions on the present topic.